\DocumentMetadata{}
\documentclass[sigconf]{acmart}

\makeatletter
\def\@ACM@checkaffil{
    \if@ACM@instpresent\else
    \ClassWarningNoLine{\@classname}{No institution present for an affiliation}%
    \fi
    \if@ACM@countrypresent\else
        \ClassWarningNoLine{\@classname}{No country present for an affiliation}%
    \fi
}
\makeatother
\usepackage{acmart-taps}

\AtBeginDocument{%
  \providecommand\BibTeX{{%
    \normalfont B\kern-0.5em{\scshape i\kern-0.25em b}\kern-0.8em\TeX}}}

\usepackage{xspace}
\usepackage{balance}

\def\ie{\textit{i.e.,}\xspace}
\def\etal{\textit{et~al.}\xspace}
\def\eg{\textit{e.g.,}\xspace}

\def\insitu{\textit{in-situ}\xspace}

\begin{document}

\copyrightyear{2026}
\acmYear{2026}
\setcopyright{cc}
\setcctype{by}
\acmConference[CHI EA '26]{Extended Abstracts of the 2026 CHI Conference on Human Factors in Computing Systems}{April 13--17, 2026}{Barcelona, Spain}
\acmBooktitle{Extended Abstracts of the 2026 CHI Conference on Human Factors in Computing Systems (CHI EA '26), April 13--17, 2026, Barcelona, Spain}
\acmDOI{10.1145/3772363.3799279}
\acmISBN{979-8-4007-2281-3/2026/04}

\author{Shaoze Zhou}
\orcid{0009-0000-3243-0599}
\email{szhou010@fiu.edu}
\affiliation{%
  \institution{Florida International University}
  \city{Miami}
  \state{FL}
  \country{USA}
}

\author{Diana Nelly Rivera Rodriguez}
\orcid{0009-0000-5369-6863}
\email{drive217@fiu.edu}
\affiliation{%
  \institution{Florida International University}
  \city{Miami}
  \state{FL}
  \country{USA}
}

\author{Pedro Remior}
\orcid{0009-0004-4648-6402}
\email{premi002@fiu.edu}
\affiliation{%
  \institution{Florida International University}
  \city{Miami}
  \state{FL}
  \country{USA}
}

\author{Joaquin Frangi}
\orcid{0009-0002-5923-3573}
\email{jfran348@fiu.edu}
\affiliation{%
  \institution{Florida International University}
  \city{Miami}
  \state{FL}
  \country{USA}
}

\author{Lingyao Li}
\orcid{0000-0001-5888-8311}
\email{lingyaol@usf.edu}
\affiliation{%
  \institution{University of South Florida}
  \city{Tampa}
  \state{FL}
  \country{USA}
}

\author{Renkai Ma}
\orcid{0000-0002-4434-2235}
\email{renkai.ma@uc.edu}
\affiliation{%
  \institution{University of Cincinnati}
  \city{Cincinnati}
  \state{OH}
  \country{USA}
}

\author{Janet G. Johnson}
\orcid{0000-0002-0456-4028}
\email{jgjanet@umich.edu}
\affiliation{%
  \institution{University of Michigan}
  \city{Ann Arbor}
  \state{MI}
  \country{USA}
}

\author{Christine Lisetti}
\orcid{0000-0003-0756-133X}
\email{lisetti@fiu.edu}
\affiliation{%
  \institution{Florida International University}
  \city{Miami}
  \state{FL}
  \country{USA}
}

\author{Chen Chen}
\orcid{0000-0001-7179-0861}
\email{chechen@fiu.edu}
\affiliation{%
  \institution{Florida International University}
  \city{Miami}
  \state{FL}
  \country{USA}
}

\ccsdesc[500]{Human-centered computing~Mixed / augmented reality}

\keywords{Proactive AI, In-Person Group Conversation, Mixed Reality, Real-Time Conversation Support}

\begin{teaserfigure}
    \centering
    \includegraphics[width=\linewidth]{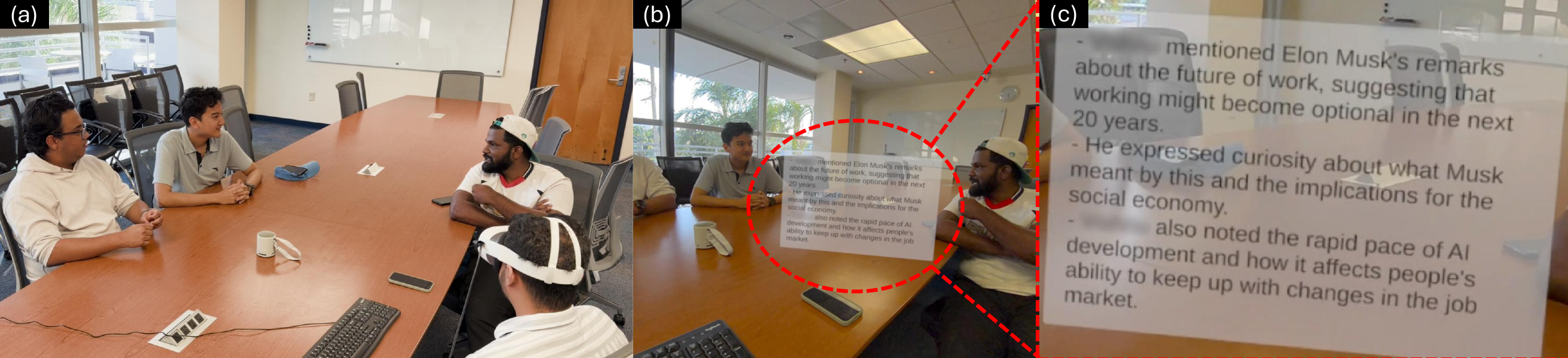}
    \caption{In-person small-group conversation experiences enabled by prototyped technology probes; (a) a third-person view showing one conversation participant wearing the MR headset; (b) a first-person view through the MR headset showing the supporting information overlay; (c) the MR supporting information showing the summary of prior conversation context.}
    \label{fig:experience}
\end{teaserfigure}

\renewcommand{\shortauthors}{Zhou~\etal}

\def\sysname{ChatMuse}

\title{Exploring Needs and Design Opportunities for Proactive Information Support in In-Person Small-Group Conversations}

\begin{abstract}
In-person small-group conversations play a crucial role in everyday life; however, facilitating effective group interaction can be challenging, as the real-time nature demands full attention, offers no opportunity for revision, and requires interpreting non-verbal cues.
Using Mixed Reality to provide proactive information support shows promise in helping individuals engage in and contribute to group conversations.
We present a preliminary participatory design and qualitative study ($N = 10$) using focus groups and two technology probes to explore the opportunities of designing proactive information support in in-person small-group conversations.
We reveal key design opportunities concerning how to maximize the benefits of proactive information support and how to effectively design such supporting information.
Our study is crucial for paving the way toward designing future proactive AI agents to enable the paradigm of \emph{augmented in-person small-group conversation experience}.

\end{abstract}

\maketitle

\balance
\section{Introduction}

In-person small-group conversations in everyday and professional settings are critical to share information, spark new ideas, build interpersonal relationships, and promote health and well-being. 
However, achieving effective in-person group conversations can be challenging and cognitively demanding, as the real-time nature requires full attention, provides no editing time, and involves understanding of non-verbal cues~\cite{Yi2021, Cooney2020, Daniel2006}. 
As a result, conversation participants need to understand complex interaction dynamics, adhere to social norms, manage multiple speakers, and tailor spoken messages for a broader audience~\cite{groupconversationchallenge}.

Recent \textbf{M}ixed \textbf{R}eality (MR) headsets show promise as conversational supporting systems by presenting relevant information through MR overlays.
In dyadic conversation scenarios, prior research has examined how critical supporting information (\eg~ keywords extracted from spoken speech~\cite{Fujimoto2025ChatAR, Jadon2024RealityChat}, relevant background context~\cite{Jadon2024RealityChat}, and guidance for subsequent utterances~\cite{Zhang2025Understood, Yang2025SocialMind}) can be used to facilitate in-person conversational experiences.
However, it remains unclear how a similar conversational supporting information can be designed or delivered for an in-person \emph{group} conversational experience. 
In particular, generalizing the design and findings of MR-based conversational support systems from dyadic to group conversations is challenging due to the increased complexity of group interactions: compared to dyadic conversations, in-person group conversations involve more complex dynamics due to the well-known `many-minds problem,' in which communication and interaction challenges increase as more participants join, making turn-taking, listening, and disclosure more difficult~\cite{Cooney2020, Sun2020}.
In a dyadic conversation, an assistant can often generate a single tailored follow-up question for the user based on the other's most recent utterance and the shared topic.
However, in small-group conversations, the same intervention may not always fit, because multiple people could speak next, speakers typically do not explicitly indicate who they are addressing, and the discussion may shift quickly between ideas.
As a result, support that works well in one-to-one interactions does not directly translate to one-to-many conversations.
Designing proactive support for a variety of conversational experiences has also shown promise in enhancing synchronous online~\cite{Liu2025, Houde2025, Leong2024} and in-person~\cite{Yang2025SocialMind} dyadic and group interactions.
\emph{Proactive AI}, rooted in the framework of mixed-initiative interaction~\cite{Allen1999, Horvitz1999}, envisions an AI agent that can autonomously determine when and how to act without requiring explicit user requests.
While prior work, \eg~SocialMind~\cite{Yang2025SocialMind}, has demonstrated the effectiveness of proactive conversational assistant systems for dyadic interactions --- where in-situ assistive MR overlays are delivered via heads-up displays on AI glasses --- it remains unclear how such proactive support can be generalized to group conversations.

We present a preliminary study~($N = 10$, two small groups with three conversation participants each and one larger group with four participants) using focus groups and technology probe~\cite{Hutchinson2003} to explore design opportunities for MR-based proactive information support in in-person small-group conversations.
Although no existing MR tool provides proactive support for in-person group conversations, we develop two experimental testbeds using Meta Quest 3S~\cite{metaquest3s}, as the technology probes, to investigate the affordances of proactive information support and the types of information users may seek when interacting with multiple conversation participants.
These two testbeds serve as inspiration for ideas generated during the participatory design process.
Our first testbed renders and simulates proactive information support through a Wizard-of-Oz~(WoZ) study~\cite{Dahlback1993}, whereas another allows users to reactively seek information support by typing during the conversation.
Figure~\ref{fig:experience} shows the experience enabled by our technology probes.
We identify key design opportunities concerning how to maximize the benefits of proactive information support and how to effectively design supporting information.
Our findings are crucial for paving ways toward designing future proactive AI agents to enable the paradigm of \emph{augmented in-person small-group conversation experience}.

\section{Methods}\label{sec::methods}

Grounded in user-centered design~\cite{Norman1996UCSD}, we aim to explore the needs and design opportunities for proactive information support in in-person group conversation experiences.
Although our participatory design~\cite{Muller1993, Bodker2018} used the Meta Quest 3S~\cite{metaquest3s}, the findings are expected to generalize to other headsets, AR glasses, and AI glasses with heads-up displays.
Our study has been approved by the Institutional Review Board.
We aim to address two \textbf{R}esearch \textbf{Q}uestions (RQs):
{\it how should proactive supporting information be designed, and in what ways can it facilitate in-person group conversation experiences? } ({\bf RQ1}); {\it what types of information are required to support in-person group conversation experiences?} ({\bf RQ2}).

\begin{figure*}[t]
    \centering
    \includegraphics[width=\linewidth]{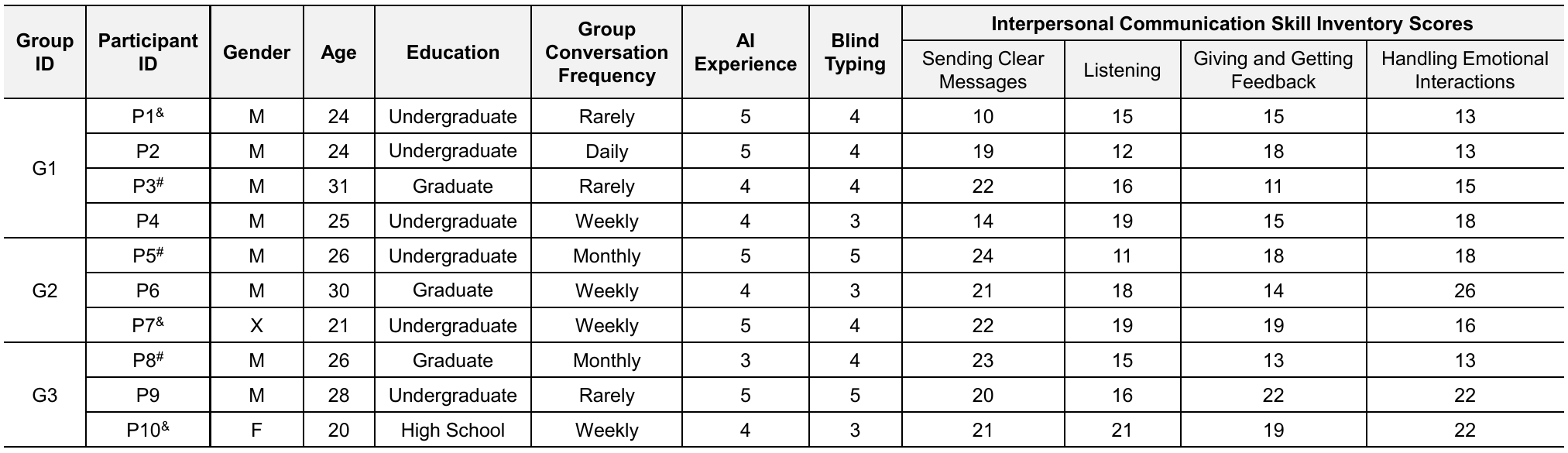}
    \vspace{-.28in}
    \caption{Participants' demographics. We used \& and \# as superscripts in participant ID to denote MR users in Phases $3$ and $4$. Responses of \emph{AI Experience} and \emph{Blind Typing} reflect participants' agreement with the given statement, {\it `I am experienced in using AI tools'} and {\it `I am experienced in typing without looking at the keyboard'} with $5$ indicating {\it strongly agree} and $1$ indicating {\it strongly disagree}. Each Interpersonal Communication Skill Inventory~\cite{Interpersonal-Communication-Skills-Inventory, Bienvenu1971, Boyd1977} score ranges from $0$ to $30$, with higher scores indicating areas of strength.}
    \vspace{-0.1in}
    \label{fig:participants}
\end{figure*}

\vspace{4px}\noindent{\bf Participants and Procedures.}
We recruited $10$~participants and formed two small groups with three participants each and one larger group with four participants (Figure~\ref{fig:participants}).
Participants were first introduced to the project and instructed to sign the informed consent.
Each group was then instructed to complete four phases.
In Phases 3 and 4, the participant group was instructed to discuss open-ended topics.
The topics included {\it where to take a Christmas trip in California} and {\it how AI may reshape the future of work}.
To prevent carry-over effects, the topics assigned to Phase $3$ and $4$ were different, with counterbalancing applied across participant groups.
Figure~\ref{fig:experience} shows an example experience enabled by technology probes prototyped in Phase 3 and 4.
All study procedures, including the MR users' first-person view, were audio- and video-recorded.
Each study session lasted approximately $90$ minutes.

\vspace{4px}\noindent$\bullet$~{\bf Phase 1: Pre-study Questionnaires.}
Participants were asked to complete demographic questionnaires, followed by the Interpersonal Communication Skills Inventory --- a standard questionnaire that assesses individuals' communication skills~\cite{Interpersonal-Communication-Skills-Inventory, Bienvenu1971, Boyd1977} (Appendix~\ref{app}).
Understanding participants' communication skills allows us to better interpret the qualitative data.

\vspace{4px}\noindent$\bullet$~{\bf Phase 2: Focus group.}
Focus groups were conducted with each participant group.
We focused on two guiding questions:
\textbf{(1)} {\it Given your prior experience, when do you think the group conversation dynamics usually deteriorate? What is your strategy to enhance the conversation experience?}
\textbf{(2)} {\it When information support can be useful during in-person group conversations, and how should such support be designed?}
Despite being guided by predefined questions, the discussions were open-ended; participants were encouraged to elaborate on their responses based on their experiences.

\vspace{4px}\noindent$\bullet$~{\bf Phase 3: Participatory design to understand the affordance of the proactive information support}. 
Although experiencing proactive information support is crucial for participants during in-person group conversations while thinking aloud about critiques and ideas, no existing tools currently provide this capability.
We therefore developed a WoZ prototype, as the technology probe, that allows conversational participants to experience proactive information support, with a researcher controlling the supporting information via a standalone dashboard.
The researcher who manually created the supporting information self-identified as being highly familiar with the conversational topics and possessing strong interpersonal communication skills.
Participants were first instructed to experience the proactive information support through MR, followed by a focus group discussion on how the supporting information could be designed.

\vspace{4px}\noindent$\bullet$~{\bf Phase 4: Participatory design to explore how conversation participants seek information during group conversation}.
While the goal of Phase 3 was to allow participants to experience proactivity, the supporting information was created by the researcher, which may have introduced unintended biases.
We included Phase 4 activities to understand what types of information are needed to support in-person small-group conversations.
To achieve this, we designed a second technology probe that allows MR users to type questions during in-person group conversations.  
To minimize disturbance from MR user's head movement while typing, only participants who can touch-type are eligible to serve as the MR user (\ie~ rate $>3$ for \emph{blind typing} in Figure~\ref{fig:participants}).
GPT-4o-mini~\cite{gpt-4o-mini} was then used to generate supporting information in response to the query input by the MR user.
An external microphone was used to capture speech, which was then transcribed and diarized using Azure's real-time transcription and speaker diarization services~\cite{AzureSpeechService}.
Participants were first invited to experience the generated supporting information, followed by a focus group discussion to share critiques, suggestions, and design ideas.

\vspace{4px}\noindent{\bf Data Analysis.}
We used thematic analysis~\cite{Braun2006} and a deductive and inductive coding approach~\cite{Lazar2010} to analyze the transcribed data. Two researchers were involved in the coding process.
Recorded videos from both third-person and first-person perspectives were referenced as the researcher read, analyzed, and coded the data.

\section{Results}
\noindent{\bf Affordance of proactive information support.}
Overall, all participants agreed that the proactive information support was helpful for maintaining focus in many conversation scenarios; yet can be unnecessary or even cause distractions in a few contexts.
A few participants also emphasized the importance and value of real-time support (\eg~{\it ``as real-time as possible would be the best!''} (P2)).
First, participants valued {\bf proactive information support when conversations stalled or drifted off-topic}. 
During the focus group of Phase 3, P1$^\&$, who scored below average in all four aspects of interpersonal communication skills, commented: {\it ``this is actually a really good thing that while having conversation, the agent is giving instructions on how to carry the conversation forward''}.
P2, one of P1$^\&$'s conversation partners, mentioned a noticeable change in participation, commenting that P1$^\&$ was {\it ``generally a bit silent,''} but appeared more engaged, possibly due to the assistance provided by the supporting information.
Participants highlighted the benefit of receiving timely, in-situ supporting information, especially when the conversation deviated from the planned topic.
Example testimonies included {\it `` [...] even when I am clueless about what to talk, the instructions were helpful in actually keeping the conversation right to the point because we were actually deviating from the topic and talking about traffic or something.''}~(P1$^\&$) and {\it ``It’s helpful for meetings in the workspace, where you have a goal and want to finish it quickly.''}~(P6).
Second, participants also appreciated {\bf receiving suggestions on how the MR user could contribute to the conversation}.
For example, P1$^\&$ suggested the usefulness of {\it``some important points which we have to consider, but we might forget it at some point''}.
As a non-MR user, P4 pointed out: {\it ``[P1$^\&$] was better at focusing on the particular topic and asking some interesting questions to make the conversations interesting.''}
P7$^\&$ highlighted its value for people who feel nervous speaking with strangers.
For example, P7$^\&$ described {\it ``I think it definitely helped [...] there's a degree of like nervousness when you're talking to strangers.''}
P5$^\#$ suggested {\it ``if there is a silence, this can help [...] sometimes you just need to get the talking going.''}
Finally, the focus group highlighted the potential benefits of {\bf providing additional information during conversations when the topic is less familiar}.
P4 highlighted that {\it ``it was about Christmas in California. We didn't know much about that. Some [supporting information] can be helpful.''}
In contrast, supporting information presented at less optimal moments could be disruptive, particularly during casual conversations.
For example, P2, who was not wearing the MR headset, described how it felt when supporting information was delivered to P1$^\&$ at less optimal moments: {\it ``I felt [P1$^\&$] came in between sometimes when we were trying to talk and be casual [...] it’s not something relevant to what we were discussing.''}

\vspace{4px}
\noindent
{\bf Supporting information must preserve, rather than replace, participants' ownership of the conversation.}
While participants recognized the value of the supporting information overlay, they stressed that it should not give the impression of replacing participants’ ownership of the conversation.
For example, P1$^\&$ noted: {\it ``I feel more comfortable and helpful [...], and rather than being pushed, I can contribute my part of the conversation with that.''}
Participants reported feeling pressured to respond when the supporting information appeared overly directive or when it did not align with their own thought processes.
Example testimonies include:
{\it ``even though I know they're just suggestions, I feel pressure to respond and relate to the instructions.''} (P7$^\&$)
and 
{\it ``It does feel like pressure [...] even if I was saying something else, I feel like I have to look at the text.''} (P10$^\&$)
P7$^\&$ further noted that perceived pressure can increase when a knowledge gap exists between the supporting information and the user (\eg~{\it ``it just said ask some activities that can be done during summer, but I don't really know how summer look like [in California].''}).
P5$^\#$ suggested: {\it ``[the supporting information] should be rendered when the conversation is going down [...] giving a suggestion every time is like creating a sense of pressure''}.

\vspace{4px}\noindent{\bf Placement of supporting information.}
Participants reported mixed opinions regarding the visual placement of the MR supporting information overlay.
A few participants considered the current design of the conversational supporting information overlay acceptable.
Example comments include: {\it ``I didn’t feel any discomfort the first time I saw it''} (P1$^\&$) and {\it ``I like to have it placed wherever I was looking [...] if it is placed on top, maybe we won't be noticing it while having a conversation. So I think it is a perfect placement [...] I think the distance was good.''} (P10$^\&$).
However, multiple key findings converged across the focus group discussions.
First, a static supporting information overlay can often cause unnecessary context switching, as users must shift attention between the conversation and the displayed information.
While P1$^\&$ acknowledged the initial placement, he also suggested: {\it ``once it appears in one location, it just stays there. If I look to the side, I can't see it.''}
P3$^\#$ emphasized the importance of placing the supporting information overlay according to the participant with whom one is conversing: {\it ``the main concern is that the [supporting information overlay] should be dynamic. It should keep moving [...] For example, if it is fixed over there [P3$^\#$ pointed to P2], then I have to take a look at him in order to read what the response is [...] if the text can move along with me, that would be nice.''}
P3$^\#$ suggested that the placement of supporting information should be adapted to the participants who are actively speaking: {\it ``if [the supporting information] can move to that person while talking, that could be really awesome [...] let's say if [P1$^\&$] is talking, [the supporting information] can move to his direction.''}
P2 and P3$^\#$ believed that, in turn, the placement of supporting information could serve as an implicit guidance for MR users regarding non-verbal cues, fostering a more engaged $f$-formation.
From a different perspective, P1$^\&$ highlighted the usefulness of adapt the placement based on physical environment: {\it``It would be nice if it could be placed on the table in front of me.''}
Second, some participants noted that the MR overlay can obstruct important non-verbal cues.
For example, P10$^\&$ described: {\it ``I think if it could adapt to the environment so it doesn’t obstruct the people in front of me.''}
P5$^\#$ suggested: {\it ``the best way would be to display text on top of the person she is speaking to [...] so you don't break the eye contact.''}

\vspace{4px}\noindent{\bf Types of supporting information.}
During Phase 4, a variety of supporting information were requested by P3$^\#$, P5$^\#$, and P8$^\#$.
Our analysis identified three types of supporting information that participants sought: summaries of the prior conversation contexts (P3$^\#$, P8$^\#$), overlooked aspects of the conversation (P8$^\#$), factual clarifications of unfamiliar topics (P5$^\#$, P8$^\#$), opinions from other conversation participants (P3$^\#$), and critical nonverbal cues from conversation partners (P7$^\&$). 
For example, P3$^\#$ described how the system's ability to summarize prior conversation contexts can be helpful: {\it ``one thing that was most helpful for me is that I was asking the agent to summarize every guest opinion [...] they were giving the summary and during that I did forget his name [refer to P4].''}
While participants believed most of the AI-inferred supporting information was useful (\eg~{\it ``overall it was good [...] the AI did recognize the name and what everybody is saying''} (P3$^\#$)), our focus groups unveiled multiple key design considerations.
P3$^\#$ suggested shortening the generated prior conversation summary to make it more readable and glanceable on-the-fly while contributing to in-person group conversation: {\it ``the summary is good, but it is a little bit longer [...] I don't know if I'm going to go down or something [P3$^\#$ refers to reading through it].''}
P7$^\&$ suggested the usefulness of indicating non-verbal cues of conversation participants because of {\it ``struggling with recognizing sometimes fairly evident social cues''}.

\section{Discussion and Conclusion}
In conclusion, our preliminary study shows the value of proactive information support for in-person small-group conversations.
Our findings suggest key design opportunities concerning how to exploit the benefits of proactive
information support, and how to effectively design such supporting information --- a key step toward designing future proactive AI agents to augment in-person small-group conversation experience.%

\vspace{4px}\noindent{\bf Design for proactivity.}
In response to {\bf RQ1}, our findings suggest that proactive support should be designed to be context-aware and aligned with moments of conversational need. Such support can facilitate in-person group conversation when participants encounter topic drift, extended silence, forgotten prior conversation context, or uncertainty about how to proceed.
A poorly designed proactive information support system can, in turn, adversely affect the group conversation experience.
Despite relying on a WoZ-based technology probe, our preliminary study identified three design opportunities for future proactive information support powered by an AI agent:
\textbf{(i)}~The system should be capable of understanding and adapting to a wide range of group conversation contexts, including both spoken speech and non-verbal cues from all participants;
\textbf{(ii)}~The \insitu~supporting information gives the MR user a sense of conversational ownership, rather than making them feel replaced.
\textbf{(iii)}~The rendering of supporting information needs to be carefully placed within the user's field of view, taking into account a variety of contexts.
These design opportunities also suggest concrete evaluation metrics for future work, such as context adaptation, conversational ownership, and the quality of different \insitu~ rendering placements.
For example, future work may explore how existing context-aware placement techniques from instructional MR \cite{Chen2023PaperToPlace, Nguyen2024PaperToPlacePatent} can be adapted and integrated to support MR-augmented in-person group conversation experience.

\vspace{4px}
\noindent{\bf Design for the supporting information.}
In response to {\bf RQ2}, our findings suggest that participants need supporting information including summaries of prior discussions, reminders of overlooked points, and reflections of individual participants' opinions. 
Our findings highlight the importance of structuring and designing such supporting information to help conversation participants interpret, coordinate, and sustain interactions.
Similar to prior work focusing on dyadic interactions~\cite{Yang2025SocialMind}, our preliminary study also suggests the usefulness of providing assistance related to non-verbal behavior, such as body gestures and voice tone.
The second critical opportunity emphasizes the importance of externalizing and structuring \insitu~information support so that it is glanceable and easy to consume.
Future work can further evaluate the quality of supporting information in terms of how easily participants can interpret and use it during in-person conversation.

\vspace{4px}
\noindent{\bf Limitations.} \emph{First}, while our primary focus is on the MR user's experience, the headset may diminish conversational partners' ability to access some MR user's non-verbal cues, \eg~gaze.
This, in turn, can affect the MR user's experience~\cite{Beyan2023}.
While most participants, acting as non-MR users, believed that the group conversation could still carry forward (\eg~{\it ``it felt a bit weird [...] but the conversation was still able to move forward''}~(P2)), future work may adopt technology prototypes with lightweight, optical see-through AR glasses that introduce less visual obstruction.
Nevertheless, given prior work that has employed similar MR headsets for co-located dyadic~\cite{Zhang2025Understood} and synchronous conversations~\cite{Johnson2025}, we believe this limitation does not diminish our contribution.
\emph{Second}, although asking participants to type queries is not natural for real-world group conversations, we used it in this exploratory study to understand the types of information participants sought while wearing the MR headset.
While participants believed they could blind type, we observed that keyboard input often disrupted the conversation flow. 
These constraints limit how well the findings reflect real-world conversations, but they also reveal concrete design opportunities for future system design.
Therefore, the findings should be interpreted as early design evidence rather than a benchmark for real-world group conversation.
\emph{Finally}, due to the nature of the preliminary study, we did not structure the study as an experiment.
The uneven group sizes reflect recruitment and scheduling constraints and may have influenced group interaction dynamics.
Consequently, we did not treat group size as an independent variable and did not make comparisons across groups.
We also acknowledge the limitations of using the WoZ prototype in Phase 3. 
Because the researcher selected and delivered proactive supporting information, the content and timing of interventions may introduce researcher bias.
Therefore, we do not treat Phase 3 findings as experimental evidence of proactive support performance, but rather as preliminary insights into the design affordances of proactivity, including when and how proactive support may be delivered in group conversations.
Future studies may involve a larger and more diverse set of participants, different group sizes, and a more rigorous comparative experiment to better understand the affordances and design of proactive supporting information for in-person small-group conversations.

\begin{acks}

We thank the anonymous reviewers for their valuable feedback. We thank Ariana Taglioretti and Jennifer Large from the Knight Foundation School of Computing and Information Sciences for their assistance with logistics during our preliminary study. We gratefully acknowledge the insights and discussions with Nirmala Arunachalam as well as senior capstone students at Florida International University during the early stages of this project's brainstorming, including Jonathan Martinez, Peace Passos, Sabrina Alvarado, Keren Rivera, and Ana Morales.

\end{acks}

\bibliographystyle{ACM-Reference-Format}
\bibliography{reference}

\appendix
\section{Phase 1 Questionnaire}\label{app}

This appendix presents the questionnaire used during Phase 1 of the study. The questions were adapted from the Interpersonal Communication Skills Inventory~\cite{Interpersonal-Communication-Skills-Inventory, Bienvenu1971, Boyd1977}. The questionnaire was used to assess participants' communication skills. It was organized into four sections, based on the structure of the original inventory: sending clear messages, listening, giving and receiving feedback, and handling emotional interactions. All questions were rated on a frequency scale of \emph{usually}, \emph{sometimes}, and \emph{seldom}. For each section, item responses were summed to produce a section-level score.

\vspace{4px}\noindent\textbf{Section I: Sending Clear Messages}

\noindent\textbf{Q1.} Is it difficult for you to talk to other people?

\noindent\textbf{Q2.} When you are trying to explain something, do others tend to put words in your mouth, or finish your sentences for you?

\noindent\textbf{Q3.} In conversation, do your words usually come out the way you would like?

\noindent\textbf{Q4.} Do you find it difficult to express your ideas when they differ from the ideas of people around you?

\noindent\textbf{Q5.} Do you assume that the other person knows what you are trying to say, and leave it to him/her to ask you questions?

\noindent\textbf{Q6.} Do others seem interested and attentive when you are talking to them?

\noindent\textbf{Q7.} When speaking, is it easy for you to recognize how others are reacting to what you are saying?

\noindent\textbf{Q8.} Do you ask the other person to tell you how she/he feels about the point you are trying to make?

\noindent\textbf{Q9.} Are you aware of how your tone of voice may affect others?

\noindent\textbf{Q10.} In conversation, do you look to talk about things of interest to both you and the other person?

\vspace{4px}\noindent\textbf{Section II: Listening}

\noindent\textbf{Q11.} In conversation, do you tend to do more talking than the other person does?

\noindent\textbf{Q12.} In conversation, do you ask the other person questions when you don't understand what they've said?

\noindent\textbf{Q13.} In conversation, do you often try to figure out what the other person is going to say before they've finished talking?

\noindent\textbf{Q14.} Do you find yourself not paying attention while in conversation with others?

\noindent\textbf{Q15.} In conversation, can you easily tell the difference between what the person is saying and how he/she may be feeling?

\noindent\textbf{Q16.} After the other person is done speaking, do you clarify what you heard them say before you offer a response?

\noindent\textbf{Q17.} In conversation, do you tend to finish sentences or supply words for the other person?

\noindent\textbf{Q18.} In conversation, do you find yourself paying most attention to facts and details, and frequently missing the emotional tone of the speakers' voice?

\noindent\textbf{Q19.} In conversation, do you let the other person finish talking before reacting to what she/he says?

\noindent\textbf{Q20.} Is it difficult for you to see things from the other person's point of view?

\vspace{4px}\noindent\textbf{Section III: Giving and Getting Feedback}

\noindent\textbf{Q21.} Is it difficult to hear or accept constructive criticism from the other person?

\noindent\textbf{Q22.} Do you refrain from saying something that you think will upset someone or make matters worse?

\noindent\textbf{Q23.} When someone hurts your feelings, do you discuss this with him/her?

\noindent\textbf{Q24.} In conversation, do you try to put yourself in the other person's shoes?

\noindent\textbf{Q25.} Do you become uneasy when someone pays you a compliment?

\noindent\textbf{Q26.} Do you find it difficult to disagree with others because you are afraid they will get angry?

\noindent\textbf{Q27.} Do you find it difficult to compliment or praise others?

\noindent\textbf{Q28.} Do others remark that you always seem to think you are right?

\noindent\textbf{Q29.} Do you find that others seem to get defensive when you disagree with their point of view?

\noindent\textbf{Q30.} Do you help others to understand you by saying how you feel?

\vspace{4px}\noindent\textbf{Section IV: Handling Emotional Interactions}

\noindent\textbf{Q31.} Do you have a tendency to change the subject when the other person's feelings enter into the discussion?

\noindent\textbf{Q32.} Does it upset you a great deal when someone disagrees with you?

\noindent\textbf{Q33.} Do you find it difficult to think clearly when you are angry with someone?

\noindent\textbf{Q34.} When a problem arises between you and another person, can you discuss it without getting angry?

\noindent\textbf{Q35.} Are you satisfied with the way you handle differences with others?

\noindent\textbf{Q36.} Do you sulk for a long time when someone upsets you?

\noindent\textbf{Q37.} Do you apologize to someone whose feelings you may have hurt?

\noindent\textbf{Q38.} Do you admit that you're wrong when you know that you are/were wrong about something?

\noindent\textbf{Q39.} Do you avoid or change the topic if someone is expressing his or her feelings in a conversation?

\noindent\textbf{Q40.} When someone becomes upset, do you find it difficult to continue the conversation?

\end{document}